**Efficiency Without Cognitive Change: Evidence from Human Interaction with Narrow AI Systems**

María Angélica Benítez[1], Rocío Candela Ceballos[1], Karina Del Valle[2], Mundo Araujo Sofía[2], Sofía Evangelina Victorio Villaroel[2], Nadia Justel[1, 2, 3]

[1] Laboratorio Interdisciplinario de Neurociencia Cognitiva (LINC), Centro de Investigación en Neurociencia y Neuropsicología (CINN), Universidad de Palermo (UP)

[2] Universidad Nacional de Córdoba (UNC)

[3] Consejo Nacional de Investigaciones Científicas y Tecnológicas (CONICET)

**Correspondence** concerning this article should be addressed to Nadia Justel Lab. Interdisciplinario de Neurociencia Cognitiva (LINC, CINN, UP). Mario Bravo 1259, CABA, Argentina. Email: nadiajustel@conicet.gov.ar

## Abstract

The rapid integration of artificial intelligence (AI) tools into educational and professional settings raises a critical applied question: does AI assistance foster durable cognitive skill development, or does it primarily facilitate situational performance? We tested this distinction using a randomized 2 (Group: AI-assisted vs. control) × 2 (Time: pre-post) mixed factorial design. Thirty adults completed standardized measures of problem solving and verbal comprehension (WAIS-III subtests) before and after a four-week intervention involving structured cognitive tasks performed either with or without access to ChatGPT. AI-assisted participants demonstrated significantly greater task-level efficiency (higher accuracy and/or faster completion across several applied tasks). However, no Group × Time interaction emerged in standardized cognitive measures, indicating no differential gains in underlying ability. These findings suggest that short-term exposure to narrow AI enhances performance efficiency without producing measurable changes in core cognitive skills. The results have implications for training, assessment, and workforce development in AI-augmented environments, highlighting the distinction between performance facilitation and skill acquisition.

# Introduction

Artificial intelligence (AI) and cognitive psychology have been intertwined since the mid-twentieth century, sharing the goal of understanding and modeling the mechanisms that underlie human thought (Newell & Simon, 1972). Originally conceived as an effort to replicate human reasoning, AI has evolved into a pervasive component of everyday life, mediating how people learn, decide, and communicate (Saxena et al., 2023). As narrow AI becomes embedded across learning, work, and communication contexts, a central applied question emerges: does AI merely facilitate performance in specific tasks, or does it produce measurable changes in underlying cognitive abilities?

Growing reliance on AI tools has intensified concerns regarding cognitive laziness, the possibility that effortless access to information reduces individuals' intrinsic motivation to engage in effortful mental activity (Grinschgl & Neubauer, 2022; Ahmad et al., 2023). Classic frameworks in information-processing and cognitive-load theory emphasize that learning and adaptive expertise depend on sustained cognitive effort and the active deployment of internal resources (Sweller, 1988; Kahneman, 2011; Vygotsky, 1962). From this perspective, external supports may optimize immediate task execution while limiting opportunities for durable skill acquisition. Narrow AI systems, in particular, are designed to optimize performance within restricted domains but show limited transfer or abstraction beyond those domains (Russell & Norvig, 2020; Tai, 2020). Understanding whether such systems alter cognitive capacity or simply redistribute cognitive labor is therefore essential for evaluating their long-term impact in educational and professional settings (Dwivedi et al., 2021).

Problem solving represents one core domain in which this distinction is especially relevant. It involves coordinated affective, behavioral, and cognitive processes that enable individuals to identify challenges, generate alternatives, and implement effective solutions (Wang & Chiew, 2010). Theoretical models differentiate between problem orientation, a motivational stance toward difficulty, and problem-solving execution, which depends on strategic reasoning and learned operational skills (Nezu & D'Zurilla, 1981; Güss et al., 2017; Ladouceur et al., 1998; Mayer, 2019). AI tools may influence both components: they can facilitate analytic reasoning and provide structured guidance, yet they may also reduce independent initiative or self-generated strategy formation (Lai et al., 2021). Empirical findings remain mixed. Some evidence suggests that dependence on AI may foster passivity or procrastination (Ahmad et al., 2023), whereas other studies indicate that AI systems can support metacognitive reflection and critical thinking when integrated into instructional contexts (Orrù et al., 2023). What remains unclear is whether short-term AI assistance produces improvements in standardized measures of problem-solving ability or whether its effects are confined to task-level performance.

Verbal comprehension constitutes a second domain in which AI assistance may exert influence. Beyond word decoding, verbal comprehension requires abstraction, semantic integration, and the interpretation of pragmatic meaning (Hauptman et al., 2022; Ness et al., 2023). Advances in natural language processing have intensified the interaction between human and artificial linguistic systems (Young et al., 2018). Contemporary language models such as ChatGPT rely on deep neural architectures capable of syntactic parsing, semantic inference, and contextual text generation (Rahman et al., 2023; Zhao, 2023). These systems have been associated with increased precision and motivation in second-language learning (Sun et al., 2023) and with personalized feedback in academic writing contexts (Mahapatra, 2024). However, whether repeated interaction with such systems modifies individuals' verbal comprehension ability, as measured through standardized assessment, remains an open empirical question.

Early adulthood provides a relevant developmental window for examining these issues. Between approximately 18 and 45 years of age, individuals typically demonstrate stable levels of advanced reasoning, flexible problem solving, and linguistic sophistication (Levinson, 1978; Arnett, 2000). This relative cognitive stability allows experimental examination of potential performance changes without the confound of rapid developmental transitions. At the same time, adults in this age range are frequent users of AI-based tools in academic and professional contexts, making the applied implications of AI-assisted cognition particularly salient.

The present study experimentally tested whether short-term interaction with narrow AI tools modifies core cognitive abilities or primarily enhances task efficiency. Using a randomized design, adults completed a series of structured cognitive activities either with or without access to ChatGPT. Raven's Progressive Matrices were administered at baseline to ensure equivalence in fluid intelligence, and selected WAIS subtests assessing problem solving and verbal comprehension were administered before and after a four-week intervention period. We formulated two hypotheses. First, we predicted that AI-assisted participants would demonstrate greater task-level efficiency, reflected in higher accuracy and/or faster completion across applied activities. Second, we predicted that AI-assisted participants would not show greater pre–post gains in standardized measures of problem-solving and verbal comprehension relative to controls. By distinguishing between performance facilitation and measurable ability change, this study seeks to clarify whether narrow AI functions as a cognitive enhancer or as an external performance scaffold within AI-augmented environments.

## Method

## Design

The study employed a randomized 2 (Group: AI-assisted vs. non-assisted control) × 2 (Time: pre-intervention vs. post-intervention) mixed factorial experimental design to evaluate the impact of narrow AI use on task performance and standardized cognitive ability. Group served as a between-subjects factor, and Time served as a within-subjects repeated-measures factor for standardized cognitive outcomes.

The primary independent variable was AI assistance during task execution, operationalized as access to ChatGPT as an external problem-solving and language-support tool during the intervention phase. The primary dependent variables were (a) task-level performance indices obtained during the programmed cognitive activities (accuracy and completion time) and (b) standardized measures of problem solving and verbal comprehension assessed via selected WAIS-III subtests administered before and after the intervention.

Raven's Progressive Matrices were administered at baseline to assess fluid intelligence and ensure initial equivalence between groups. RPM scores were not treated as outcome variables but as baseline indicators of general cognitive ability.

Participants were randomly assigned to either the AI-assisted condition or the non-assisted control condition. The design permitted examination of (a) between-group differences in task-level performance and (b) Group × Time interaction effects in standardized cognitive measures, thereby testing whether exposure to narrow AI produces measurable changes in underlying ability or primarily enhances situational task efficiency.

The study adhered to established methodological standards for experimental research in behavioral sciences (Hernández Sampieri et al., 2006). All sessions were conducted online under controlled and standardized conditions. Procedures were preregistered and implemented in accordance with the ethical and methodological guidelines of the American Psychological Association (APA).

**Participants**

Participants were recruited through an open online call distributed via social media platforms (WhatsApp and Instagram) and via snowball sampling. Eligible individuals were adults aged 18 to 45 years with no prior formal training in artificial intelligence or computational fields. This age range was selected to examine AI-assisted cognition within a cognitively stable adult population, minimizing developmental variability in reasoning and verbal abilities.

Exclusion criteria included a self-reported history of neurological, psychiatric, or developmental disorders, as well as any medical condition that could interfere with task performance.

The final sample consisted of 30 participants (aged 18-45), randomly assigned to the AI-assisted group (n = 15) or the non-assisted control group (n = 15). Random assignment was conducted prior to baseline assessment. Groups did not differ significantly in age, years of education, or self-reported technology familiarity, confirming initial equivalence at baseline.

Participants in the AI-assisted condition were granted access to ChatGPT during the intervention tasks, whereas participants in the control condition completed the same activities without any AI assistance. No participants were excluded after randomization, and all completed the full study protocol.

All participants provided informed consent prior to participation and took part voluntarily without financial compensation. The study was conducted in accordance with the Declaration of Helsinki and the ethical standards of the American Psychological Association (APA, 2017).

## Instruments and activities

### *Instruments*

All standardized assessments and programmed activities were administered online via synchronous screen sharing. Item order, instructions, and time constraints were maintained in accordance with each test manual to preserve standardization.

### Sociodemographic Questionnaire

A custom questionnaire (Google Forms) was used to collect demographic and contextual information, including age, gender, handedness, city and nationality, educational level, socioeconomic status, and prior experience with AI tools. Technology familiarity was assessed through self-report.

### Raven's Progressive Matrices (RPM; Raven et al., 2003)

The RPM was administered at baseline as a measure of fluid intelligence and abstract reasoning to ensure comparability between groups prior to the intervention. The instrument consists of 60 nonverbal items organized into five series (A–E) of increasing difficulty. Participants selected the missing element that completed each visual pattern. RPM scores were used solely as baseline indicators of general cognitive ability and were not treated as outcome variables.

### Wechsler Adult Intelligence Scale – Third Edition (WAIS-III; Wechsler, 1997)

Four WAIS-III subtests were administered before and after the intervention to assess standardized indices of problem-solving and verbal comprehension ability:

1. **Picture Completion** assessed perceptual organization and visual reasoning by requiring identification of missing elements in incomplete drawings under time constraints (approximately 20 s per item).
2. **Arithmetic** assessed quantitative reasoning and working memory through orally presented arithmetic problems solved without external aids under time limits.

3. **Similarities** measured abstract verbal reasoning by requiring participants to identify conceptual relationships between paired words.
4. **Vocabulary** assessed lexical knowledge and verbal expression through definitions of words of increasing difficulty.

Standard administration and scoring procedures outlined in the WAIS-III manual were followed. Raw scores were computed according to manual guidelines and used as standardized indicators of underlying cognitive ability. The full assessment battery required approximately 45-60 minutes to complete.

***Programmed Activities***

Eight structured cognitive activities were developed for the intervention phase to engage problem-solving and verbal-comprehension processes under both AI-assisted and non-assisted conditions. These activities served as performance tasks during the intervention and were not designed as standardized measures of cognitive ability.

All tasks were implemented using Canva templates and administered synchronously via Google Meet with screen sharing to ensure standardized presentation. Participants in the AI-assisted condition were granted unrestricted access to ChatGPT during task completion, whereas participants in the control condition completed the same activities without external assistance.

The intervention spanned four consecutive weeks, with two sessions per week (approximately 30-60 minutes per session). Task design was informed by prior research in educational psychology, cognitive training, and problem-solving instruction (e.g., Baddeley, 2003; Beaty et al., 2015; De Bono, 1970; Guilford, 1967; Güss et al., 2017; Karpicke & Blunt, 2011; Kintsch, 1998; Mayer, 2019; Nation, 2013; Perfetti & Stafura, 2014; Roediger & Karpicke, 2006; Runco, 2010; Simon, 1975; van den Broek & Helder, 2017). Task materials are available as supplementary material (Benítez et al.,

2025).

The activities included:

1. **Crossword Puzzle.** A thematic crossword targeting verbal comprehension and lexical retrieval through semantic mapping strategies (Baddeley, 2003; Nation, 2013).
2. **Problem-Solving Tasks.** Arithmetic and logical reasoning problems emphasizing analytical reasoning and strategic planning (Mayer, 2019; Güss et al., 2017).
3. **Word-Guessing Game.** Lexical inference task engaging semantic integration processes (Perfetti & Stafura, 2014).
4. **Lateral Thinking.** Riddle-based problems requiring non-linear reasoning and cognitive flexibility (De Bono, 1970; Runco, 2010).
5. **Trivia.** Timed multiple-choice questions assessing retrieval and decision speed (Karpicke & Blunt, 2011; Roediger & Karpicke, 2006).
6. **Tower of Hanoi.** Planning and executive sequencing task with progressive complexity (Simon, 1975).
7. **Reading Comprehension.** Short expository text followed by comprehension questions targeting inferential understanding (Kintsch, 1998; van den Broek & Helder, 2017).
8. **Brainstorming.** Timed idea-generation task promoting divergent thinking and fluency (Guilford, 1967; Beaty et al., 2015).

Accuracy and/or completion time were recorded for each task as indices of situational performance efficiency.

Critically, these activities were analyzed exclusively as intervention-phase performance measures. They were not intended to serve as indicators of latent cognitive ability. Core cognitive abilities were assessed solely through standardized neuropsychological instruments (RPM and WAIS-III subtests) administered outside the intervention context. By employing distinct item formats, response demands, and measurement frameworks across intervention tasks and standardized assessments, the study minimized criterion contamination and preserved construct validity between performance facilitation and ability measurement.

## Procedure

All data were collected online via Google Meet using synchronous screen sharing to ensure standardized visual presentation of materials. Stimuli were displayed through PowerPoint and Canva. All sessions were conducted by the same team of three trained experimenters following standardized scripts and administration protocols to minimize procedural variability.

The study followed a structured seven-week protocol.

### *Week 1 (Enrollment and Demographics)*

Participants completed informed consent and the sociodemographic questionnaire via Google Forms. The consent form outlined study objectives, confidentiality procedures, voluntary participation, and the right to withdraw at any time without penalty.

### *Week 2 (Baseline Assessment)*

Participants completed Raven's Progressive Matrices (RPM) to assess fluid intelligence and the selected WAIS-III subtests (Arithmetic, Picture Completion, Similarities, and Vocabulary) to establish pre-intervention measures of problem-solving and verbal comprehension ability.

### *Weeks 3–6 (Intervention Phase)*

Participants completed eight programmed cognitive activities distributed across four consecutive weeks (two sessions per week, with at least one rest day between sessions to minimize fatigue). Each session lasted approximately 30–60 minutes, depending on task complexity. Tasks included Crossword Puzzle, Problem-Solving Exercises, Word-Guessing Game, Lateral Thinking, Trivia, Tower of Hanoi, Reading Comprehension, and Brainstorming.

Participants in the AI-assisted condition were granted unrestricted access to ChatGPT during task completion. No prompts were pre-specified, monitored, or recorded, and participants were free to use the tool at their discretion. Participants in the control condition completed identical tasks without AI assistance. This design preserved experimental contrast while maintaining ecological validity.

***Week 7 (Post-Intervention Assessment)***

Participants completed the same WAIS-III subtests administered at baseline, enabling pre–post comparisons of standardized cognitive measures.

All participants completed the full seven-week protocol, and no attrition occurred after randomization.

**Ethical Considerations**

This study was conducted in full compliance with Argentina’s National Law No. 25.326 on the Protection of Personal Data, National Mental Health Law No. 26.657, the Declaration of Helsinki (World Medical Association, 2013), and the Ethical Principles of Psychologists and Code of Conduct of the American Psychological Association (American Psychological Association, 2017). The study protocol received prior approval from the institutional ethics committee.

Participation was voluntary, and all participants provided informed consent before enrollment. Participants were informed of their right to withdraw at any time without penalty. Confidentiality and anonymity were ensured through secure data storage and analysis of aggregated data only.

Upon completion of the study, participants were debriefed and offered a general summary of the findings and a description of the experimental activities, without disclosure of individual performance outcomes.

**Data analysis**

Sociodemographic variables (age, years of education, and prior experience with AI) were analyzed descriptively to characterize the sample. Baseline equivalence between groups was examined using independent-samples t-tests (or non-parametric equivalents when necessary). Raven's Progressive Matrices (RPM) scores were also compared between groups at baseline using independent-samples t-tests to ensure equivalence in fluid intelligence prior to the intervention.

To evaluate the impact of AI assistance on standardized cognitive measures, separate 2 (Group: AI-assisted vs. control) × 2 (Time: pre-intervention vs. post-intervention) mixed factorial analyses of variance (ANOVAs) were conducted for each WAIS-III subtest. Group was treated as a between-subjects factor and Time as a within-subjects factor. The critical test of differential cognitive change was the Group × Time interaction. Significant main effects and interactions were followed by planned comparisons when appropriate.

Performance during the programmed intervention activities was analyzed as situational task efficiency. For each task, accuracy and/or completion time were compared between groups using independent-samples t-tests. When relevant, effect sizes (Cohen's d) were calculated to quantify the magnitude of group differences.

Assumptions of normality and homogeneity of variances were assessed using Shapiro–Wilk (or Kolmogorov–Smirnov) and Levene's tests, respectively. When assumptions were violated, appropriate non-parametric alternatives (e.g., Mann–Whitney U or Wilcoxon tests) were applied. All analyses were conducted with a significance level of $\alpha = .05$. Statistical analyses were performed using IBM SPSS Statistics (Version 26).

## Results

Analyses were conducted on the full sample ($N = 30$; $n = 15$ AI-assisted, $n = 15$ control). Assumptions of normality and homogeneity of variances were evaluated using Shapiro–Wilk and Levene's tests, respectively. For most variables, assumptions were met ($p > .05$), and non-parametric analyses yielded convergent results.

### Sociodemographic Characteristics and Baseline Equivalence

Baseline equivalence analyses were conducted to verify comparability between the AI-assisted and control groups prior to the intervention.

No statistically significant differences were observed in age, $t(29) = 2.03$, $p = .052$, years of education, $t(29) = -0.24$, $p = .813$, or years of prior AI experience, $t(29) = -0.89$, $p = .380$. Additionally, the distribution of sex did not differ significantly between groups, $\chi^2(1) = 0.58$, $p = .446$.

Descriptively, participants in the control group were slightly older ($M = 26.63$, $SD = 6.71$) than those in the AI-assisted group ($M = 22.53$, $SD = 4.10$), although this difference did not reach statistical significance. Years of education were comparable across groups (Control: $M = 13.50$, $SD = 1.26$; Experimental: $M = 13.60$, $SD = 1.06$), as was prior AI experience, which was minimal in both conditions.

Baseline fluid intelligence, measured through Raven's Progressive Matrices, also did not differ significantly between groups, $t(29) = -1.14$, $p = .263$ (Control: $M = 40.94$, $SD = 10.24$; Experimental: $M = 44.60$, $SD = 7.26$).

Together, these findings support initial comparability between groups prior to the intervention.

**Standardized Cognitive Measures (WAIS-III Subtests)**

To evaluate the impact of AI assistance on standardized cognitive measures, separate 2 (Group: AI-assisted vs. control) × 2 (Time: pre-intervention vs. post-intervention) mixed ANOVAs were conducted for each WAIS-III subtest. No significant Group × Time interactions were observed for Picture Completion, $F(1, 29) = 0.30$, $p = .587$, $\eta_p^2 = .010$, Arithmetic, $F(1, 29) = 0.50$, $p = .483$, $\eta_p^2 = .017$, Vocabulary, $F(1, 29) = 0.11$, $p = .743$, $\eta_p^2 = .004$, or Similarities, $F(1, 29) = 0.70$, $p = .408$, $\eta_p^2 = .024$, indicating that pre–post changes did not differ between groups. A significant main effect of Time was observed for Picture Completion, $F(1, 30) = 18.75$, $p < .001$, $\eta_p^2 = .385$, and Arithmetic, $F(1, 30) = 15.42$, $p < .001$, $\eta_p^2 = .340$, reflecting overall improvement across both groups. A main effect of Time was also observed for Vocabulary, $F(1, 30) = 5.93$, $p = .021$, $\eta_p^2 = .165$, indicating a decline in scores from pre to post across both groups. No significant main effects of Group were observed (all $p > .05$).

**Figure 1**

*Picture Completion performance in the Control (A) and AI-assisted (B) groups*

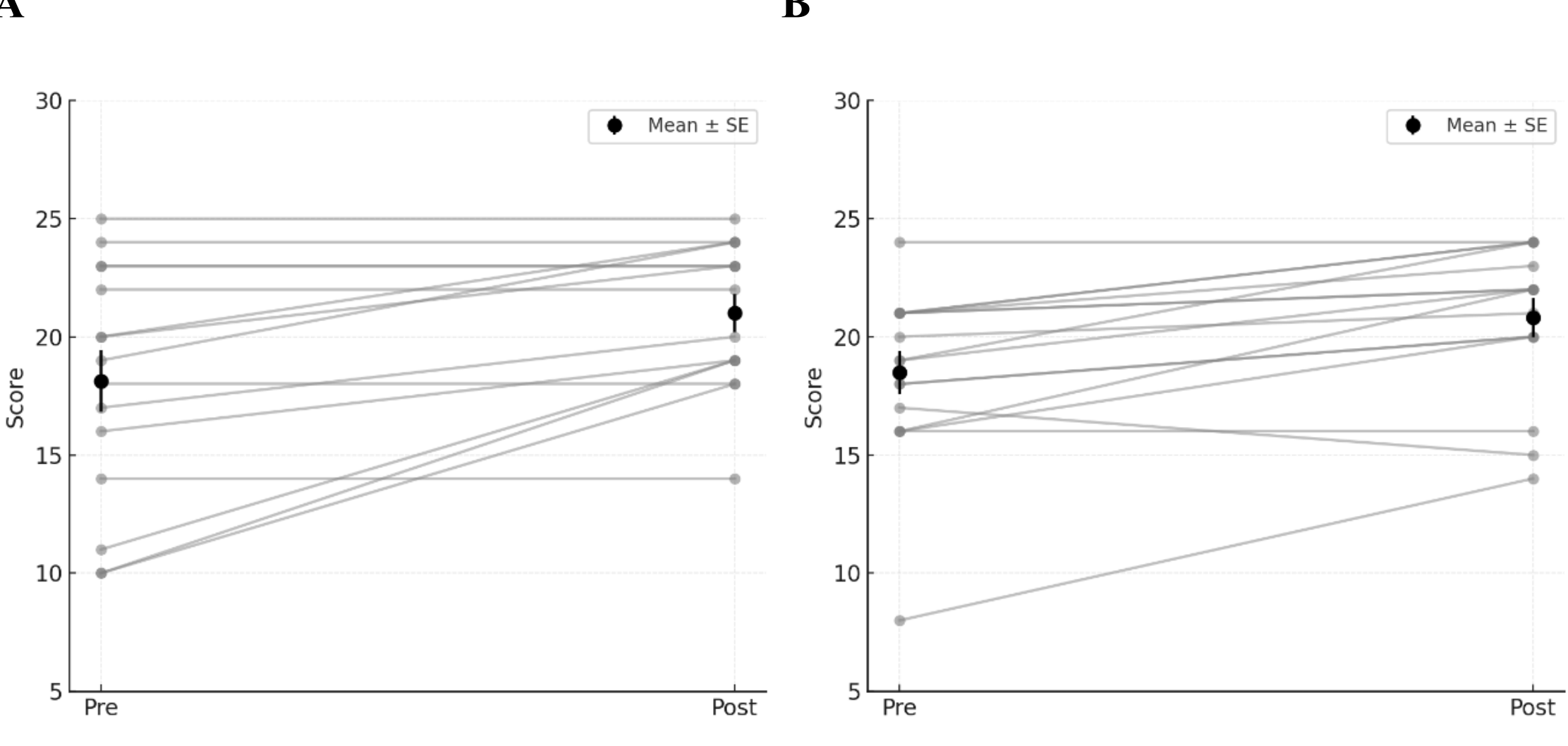

*Note.* Each line represents an individual participant's pre-post scores. Black dots indicate group means ± SE.

**Figure 2**

*Arithmetic performance in the Control (A) and AI-assisted (B) groups*

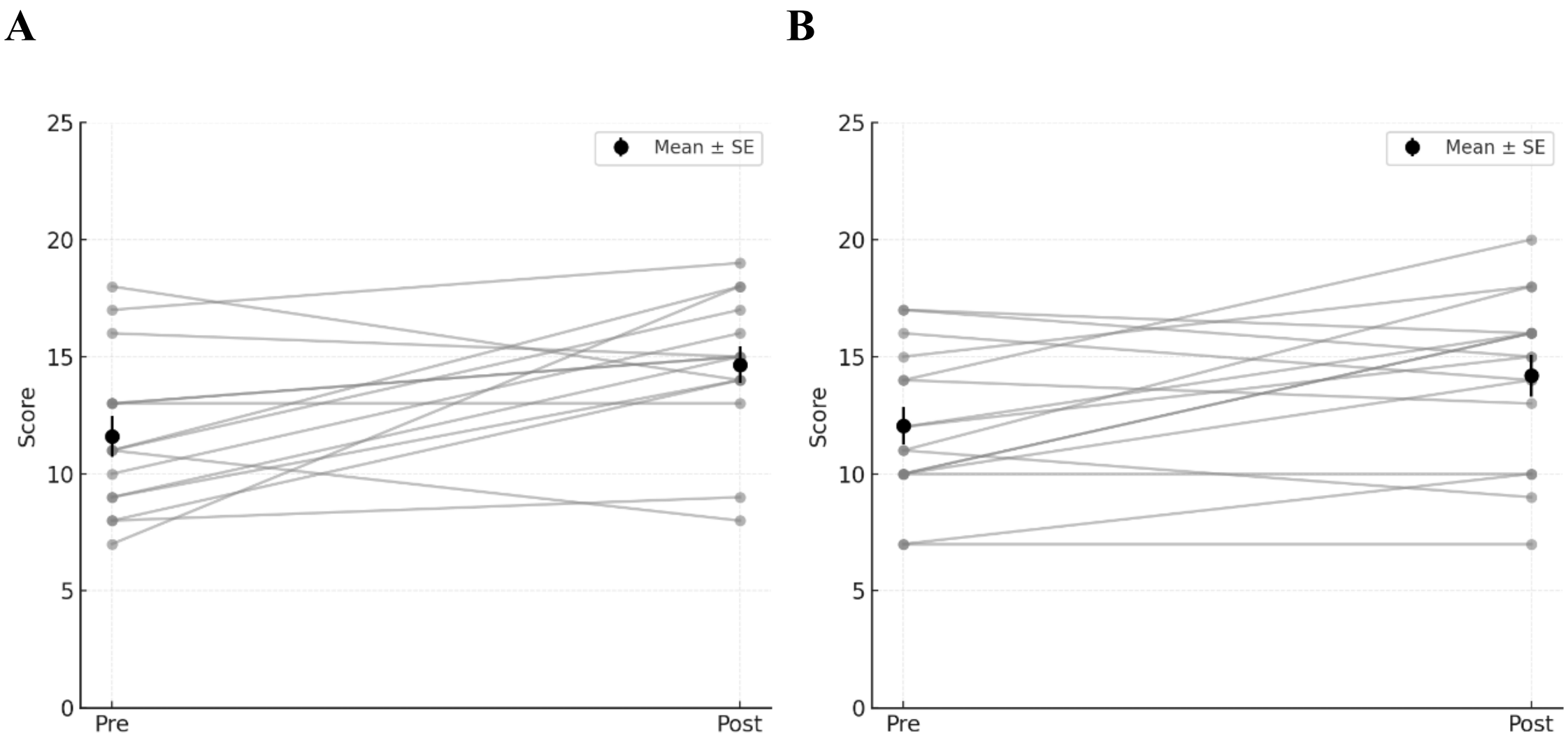

*Note.* Each line represents an individual participant's pre–post scores. Black dots indicate group means ± SE.

## Performance in Programmed Activities

Independent-samples t-tests were conducted to compare situational task efficiency between the AI-assisted and control groups across the programmed activities.

### *Crossword*

AI-assisted participants achieved higher accuracy than controls ($t(29) = -5.67$, $p < .001$, $d = 1.25$; see Figure 3).

**Figure 3**

*Crossword task performance in the Control and AI-assisted groups*

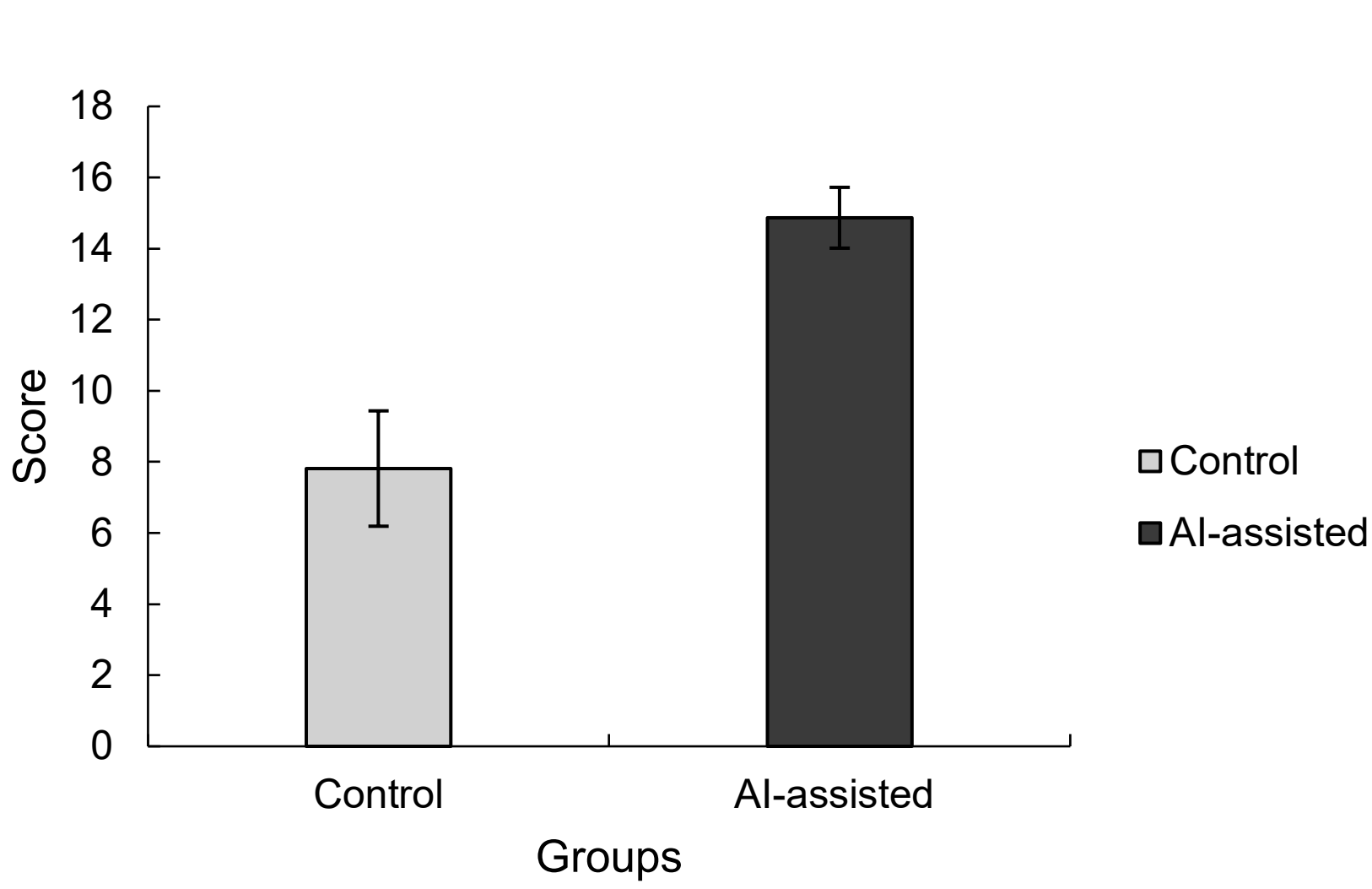

*Note.* Mean accuracy scores in the Crossword task for the Control and AI-assisted groups (**$p < .001$). Error bars represent standard errors of the mean.

### ***Problem-Solving Task***

Performance was also superior in the AI-assisted group ($t(29) = -4.32$, $p < .001$, $d = 0.98$; see Figure 4).

**Figure 4**

*Problem-Solving task performance in the Control and AI-assisted groups*

**

4.5
4
3.5
3
2.5
2
1.5
1
0.5
0

Score

Control
AI-assisted

Groups

Control
AI-assisted

*Note.* Mean accuracy scores in the Problem-Solving task for the Control and AI-assisted groups (**$p < .001$). Error bars represent standard errors of the mean.

### *Word-Guessing Game*

AI assistance improved both accuracy ($t(29) = -2.44$, $p = .021$, $d = 0.58$) and response time ($t(29) = 3.48$, $p = .002$, $d = 0.80$; see Figures 5–6).

**Figure 5**

*Word-Guessing Game task performance in the Control and AI-assisted groups*

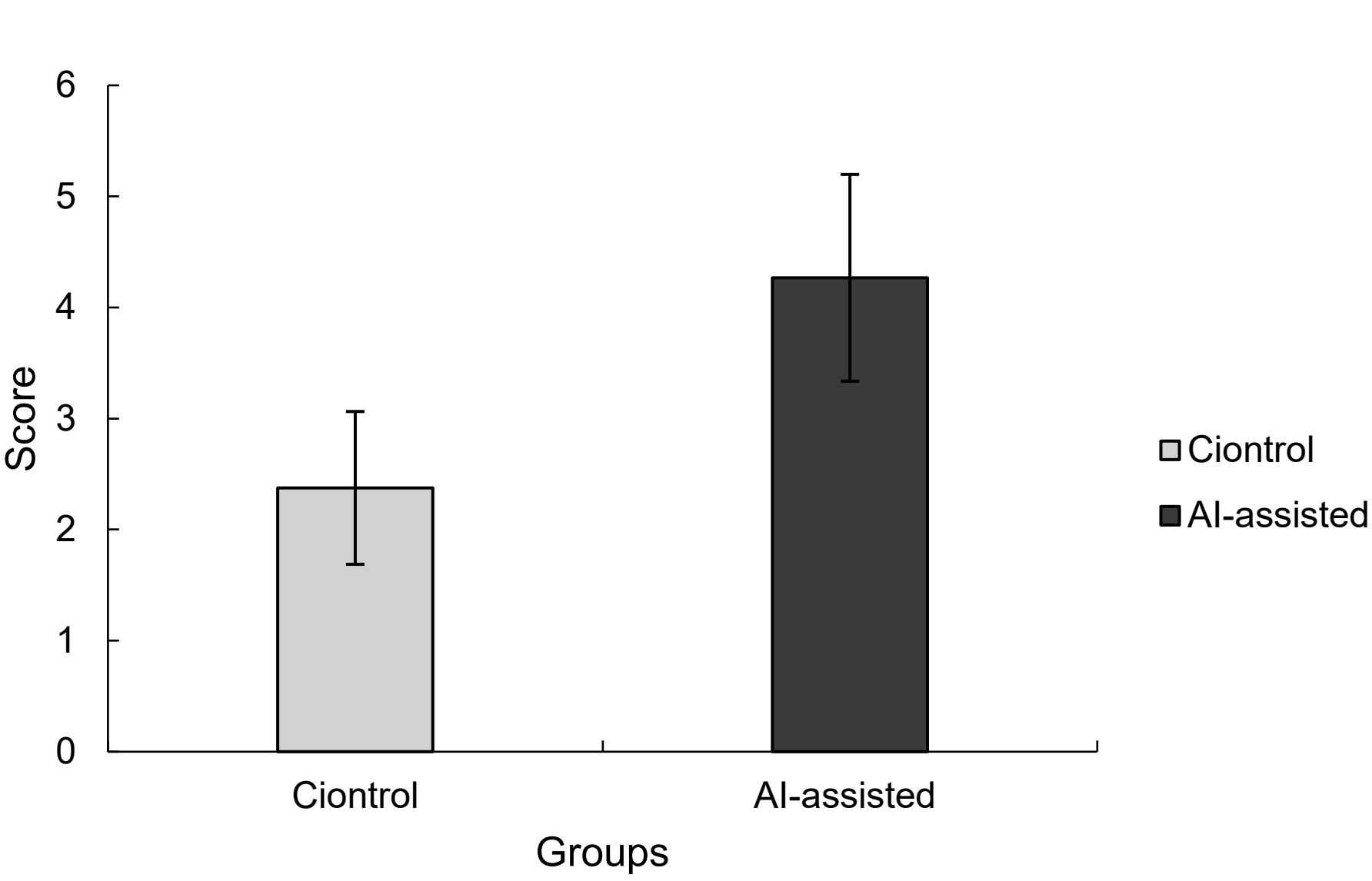

*Note.* Mean accuracy scores in the Word-Guessing Game task for the Control and AI-assisted groups (**$p$ < .05). Error bars represent standard errors of the mean.

**Figure 6**

*Word-Guessing Game task time performance in the Control and AI-assisted groups*

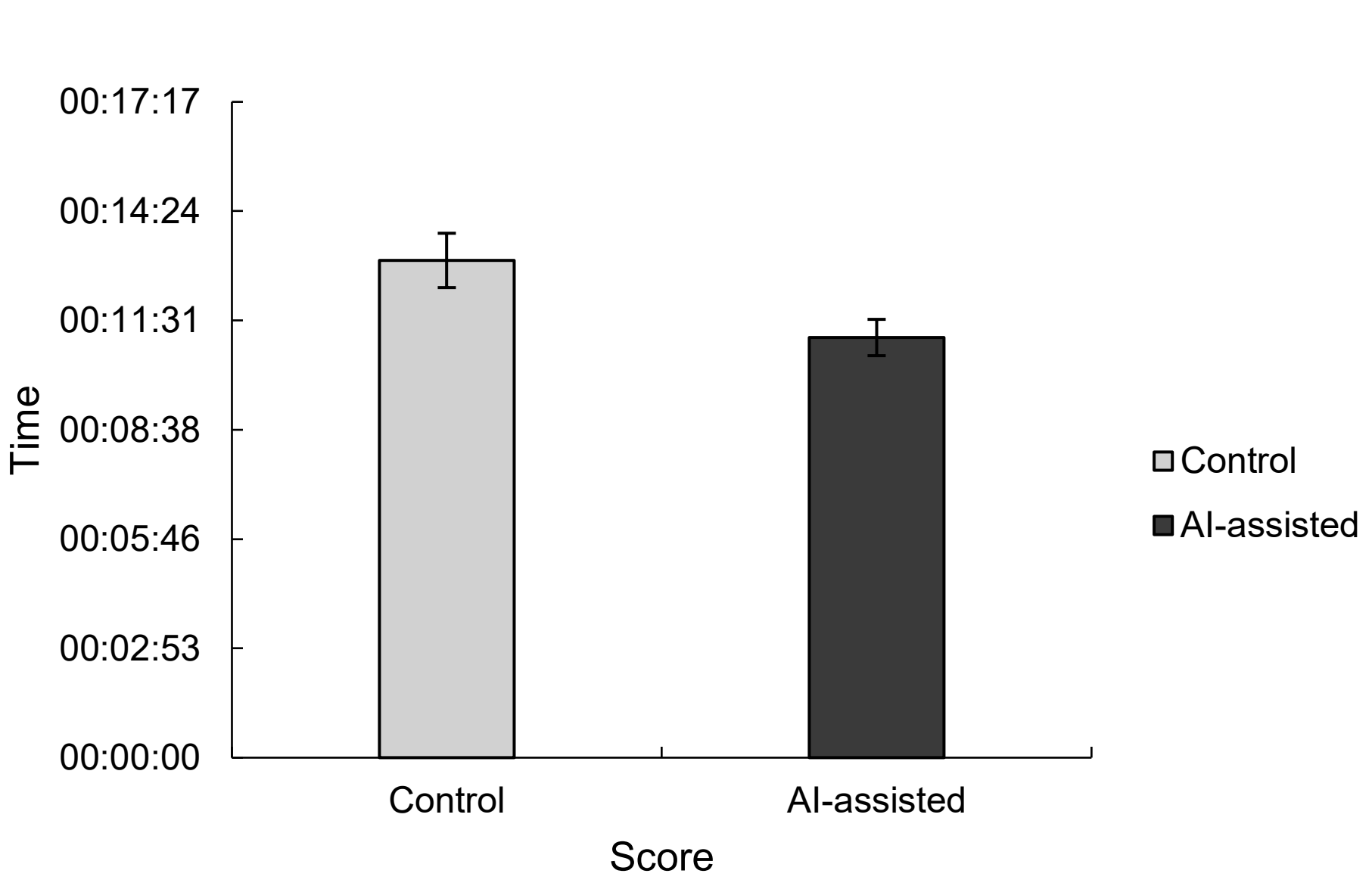

*Note.* Mean accuracy scores in the Word-Guessing Game time task for the Control and AI-assisted groups (**$p$ < .05). Error bars represent standard errors of the mean.

### *Trivia*

Again, the AI-assisted group outperformed controls ($t(29) = -6.57$, $p < .001$, $d = 1.45$; see Figure 7).

**Figure 7**

*Trivia task performance in the Control and AI-assisted groups*

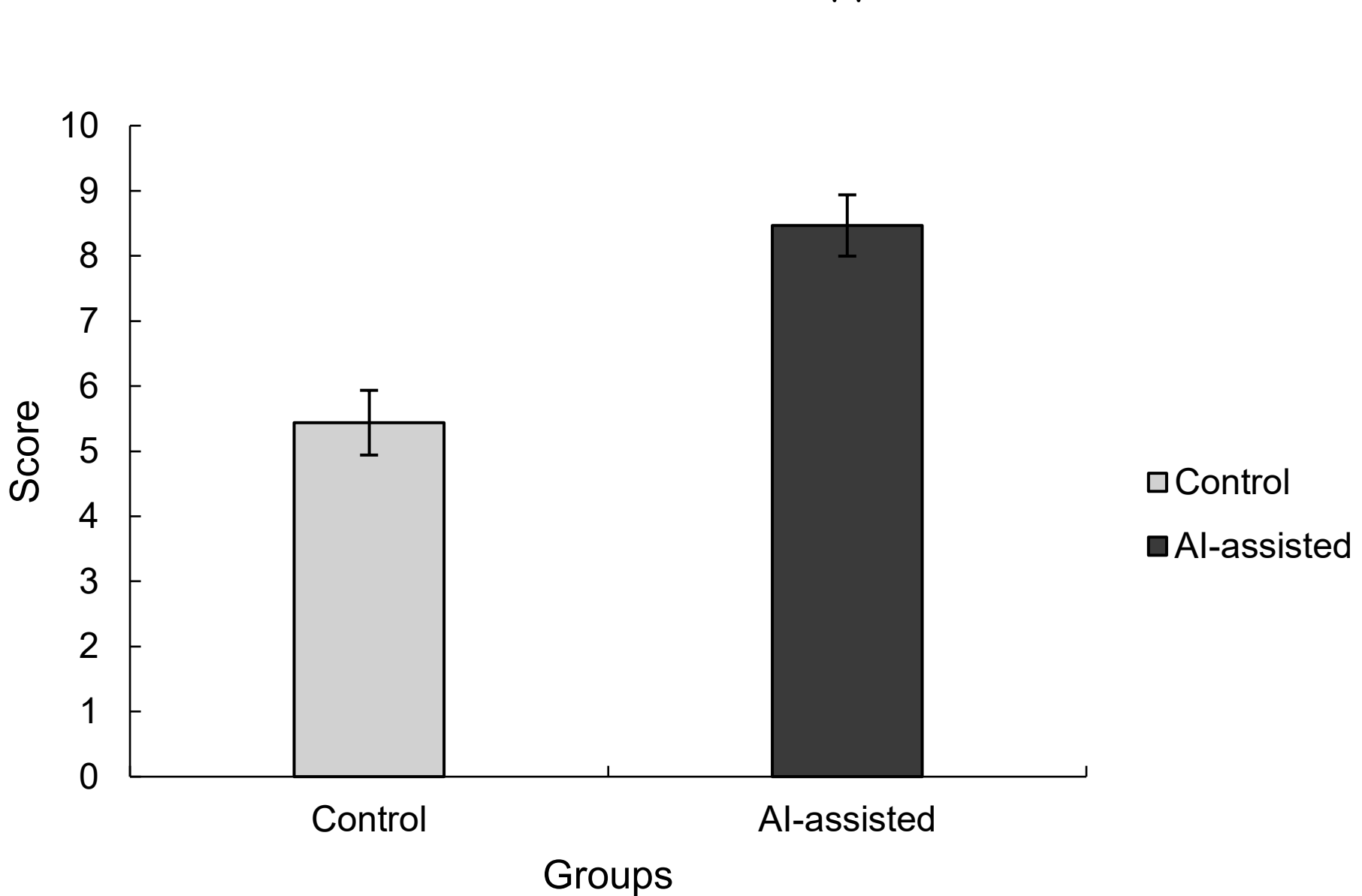

*Note.* Mean accuracy scores in the Trivia task for the Control and AI-assisted groups (**$p < .001$). Error bars represent standard errors of the mean.

### *Tower of Hanoi*

A trend toward faster completion time was observed in the AI-assisted group ($t(29) = 2.03$, $p = .051$, $d = 0.46$; see Figure 8).

**Figure 8**

*Tower of Hanoi task time performance in the Control and AI-assisted groups*

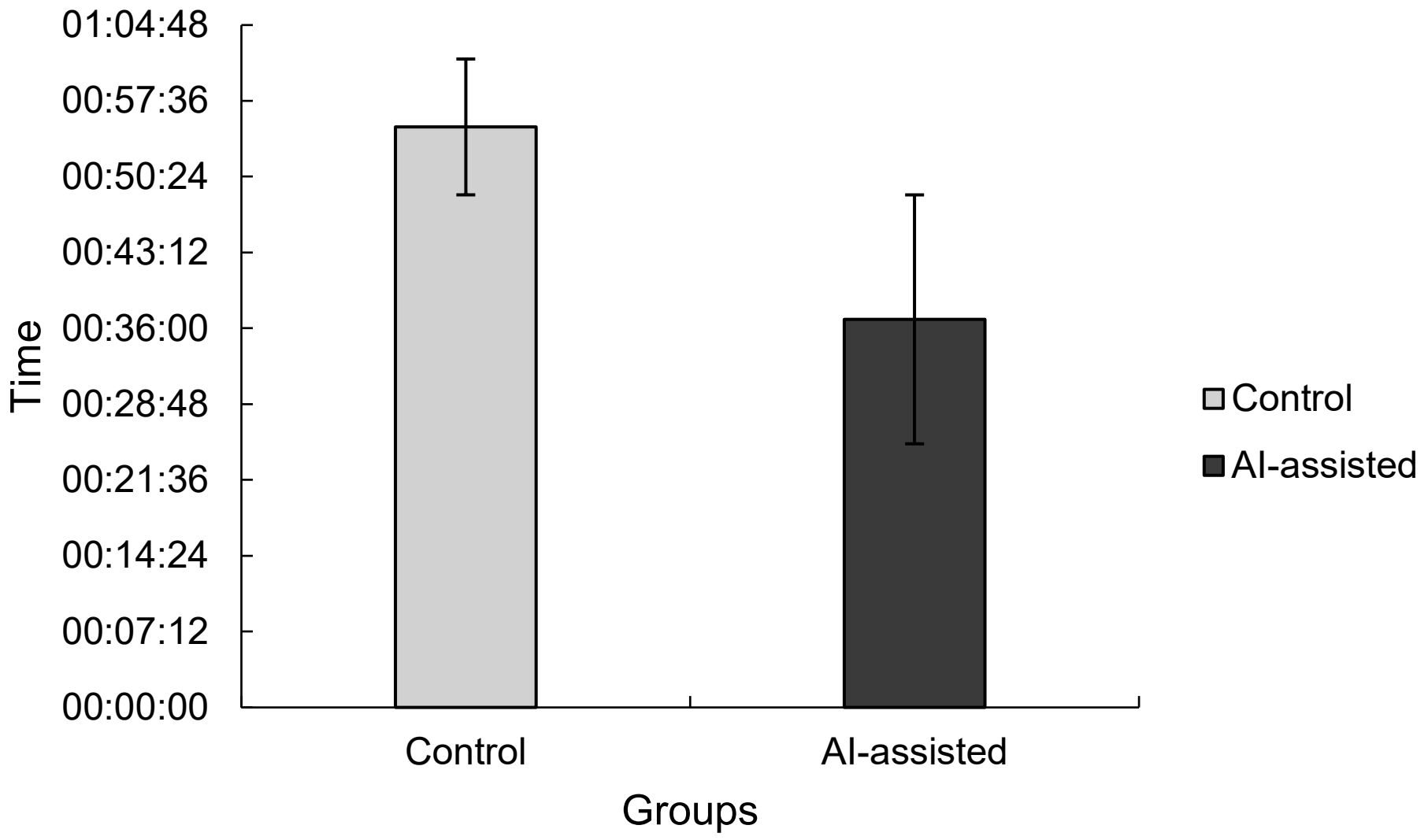

*Note.* Mean accuracy scores in the Tower of Hanoi for the Control and AI-assisted groups. Error bars represent standard errors of the mean.

No significant differences emerged between groups in the Lateral Thinking, Tower of Hanoi and Brainstorming activities or in completion times for the Crossword, Problem-Solving, Lateral Thinking, Trivia, or Brainstorming tasks ($p > .05$).

Overall, these findings demonstrate that AI assistance consistently enhanced efficiency and accuracy across several applied cognitive tasks, with medium to large effect sizes, but did not produce measurable changes in standardized cognitive abilities. These results support the hypothesis of efficiency gains without cognitive change, suggesting that narrow AI primarily serves as an external performance optimizer rather than a driver of intrinsic cognitive enhancement.

## Discussion

The present study examined whether short-term interaction with narrow AI tools influences verbal comprehension and problem-solving abilities in adults aged 18 to 45 years. As predicted, exposure to AI-assisted tasks improved performance efficiency but did not produce measurable changes in standardized indices of these abilities, as assessed by neuropsychological instruments (Raven et al., 2003; Wechsler, 1997). In other words, brief engagement with narrow AI optimized how participants performed tasks without modifying the underlying cognitive architecture associated with reasoning and language comprehension.

This finding aligns with previous research on the cognitive consequences of digital technologies, which suggests that such tools may facilitate performance in task-specific contexts without enhancing the cognitive mechanisms that support those performances (Bainbridge & Mayer, 2018). Given that crystallized abilities require sustained engagement to show measurable gains (Klingberg, 2010), the effects observed here reflect process optimization rather than cognitive restructuring. Cognitive change, in this sense, may demand prolonged and intensive interaction, conditions not replicated in short-duration interventions.

Participants in the AI-assisted group consistently displayed greater efficiency and faster completion times in applied cognitive tasks. This dissociation between efficiency and ability supports the framework of extended cognition, which posits that external artifacts can temporarily expand cognitive capacity by offloading low-level operations (Kirsh, 2010). From this perspective, AI functions as a cognitive scaffold that reduces processing demands and allows individuals to allocate resources to higher-order reasoning (Sweller et al., 2011; Mayer, 2019). According to Cognitive Load Theory, performance in complex tasks improves when working-memory resources are liberated from routine operations (Sweller, 1988; Van Merriënboer & Sweller, 2005). In the present study, AI-assisted participants benefited from such redistribution of cognitive effort, achieving superior efficiency without corresponding gains in cognitive ability.

The absence of measurable cognitive improvement despite enhanced efficiency invites a deeper theoretical reflection. Fisher et al. (2015) proposed that cognitive tools, when over-relied upon, may stabilize rather than develop internal competences, particularly when not integrated into metacognitively guided learning contexts. This notion resonates with *offloading* theories (Risko & Gilbert, 2016), which argue that while delegating cognitive work to external systems optimizes performance, it may simultaneously reduce the frequency and depth of self-generated reasoning cycles. Hence, AI may promote a form of functional intelligence, highly adaptive within structured problem spaces, but not necessarily a growth in autonomous cognitive control.

Improvements observed between pre- and post-intervention assessments across both groups likely reflect practice and familiarity effects (Ericsson, 2016). Deliberate Practice Theory posits that performance enhancement depends on intentional, feedback-driven repetition over extended periods (Macnamara et al., 2014). The short duration of this intervention thus represents a boundary condition: longitudinal research indicates that measurable changes in executive and reasoning functions typically require prolonged or adaptive training (Jaeggi et al., 2008; Karbach & Verhaeghen, 2014). Future studies should therefore employ extended or adaptive designs capable of distinguishing transient performance facilitation from durable cognitive restructuring. From a critical standpoint, these findings raise questions about cognitive complacency in technologically saturated environments. Theories of cognitive idleness suggest that effortless access to algorithmic solutions can attenuate intrinsic motivation and diminish engagement in effortful thought (Sparrow et al., 2011; Carr, 2011). Such dynamics risk transforming not cognition itself, but the ecology of cognitive effort, how and when humans choose to think. The challenge is therefore both empirical and ethical: integrating AI as an augmentative resource without eroding self-regulatory and metacognitive engagement.

Our findings, while informative, are subject to several important methodological boundary conditions that inform the next steps in this research field. First, the modest sample size ($N = 30$) and non-probabilistic recruitment characterize this work as an essential, pilot study. While we observed medium-to-large effect sizes (*d* up to 1.45) demonstrating enhanced efficiency, the current sample limits statistical power to detect subtle, yet potentially durable, changes in standardized cognitive abilities. Future research necessitates a multi-site, fully randomized controlled trial with a larger sample to validate the "efficiency without cognitive change" hypothesis with greater certainty and generalizability.

Second, our reliance solely on standardized behavioral neuropsychological instruments (WAIS-III) for measuring cognitive ability limits the depth of our conclusions regarding the underlying cognitive mechanisms. The claim that AI acts as a cognitive scaffold, offloading effort without modifying internal architecture, is a theoretical inference. Longitudinal and neurocognitive studies are urgently needed. These should integrate neuroimaging techniques (fMRI or EEG) to track changes in neural plasticity and the cerebral distribution of cognitive labor. Furthermore, the inclusion of dual-task paradigms or high-resolution attentional control measures could precisely quantify the reduction in cognitive load afforded by the AI system.

Third, our ecological approach, which granted participants unrestricted access to ChatGPT, precluded the systematic analysis of the human–AI interaction patterns. We did not monitor or record the specific prompts used, which is a significant missed opportunity for mechanistic insight. Was efficiency gained by offloading low-level search operations or by using the AI for metacognitive validation and structuring of the problem? Future studies must incorporate real-time capture and content analysis of the AI prompts to distinguish between shallow offloading (using the AI for answers) and reflective offloading (using the AI for strategy).

Finally, the design resulted in significant differences in time-on-task between groups, potentially confounding exposure effects. Subsequent interventions should employ an adaptive design that statistically controls for the duration of engagement or equates the total active thinking time spent on the tasks, ensuring that both groups receive comparable mental stimulation.

In conclusion, these findings provide a initial evidence for ethical and educational policy. Our results suggest the need for scaled, neurocognitive, and process-oriented research to fully map the evolving cognitive ecology shaped by narrow AI.

**Conclusions**

This study contributes to the growing body of evidence on how narrow AI tools interact with human cognition. Results demonstrate that while AI enhances performance efficiency and speed by reducing cognitive load, these benefits do not translate into measurable improvements in core cognitive abilities over short periods. In other words, AI optimizes how we perform, not how we think.

From a psychological and educational perspective, AI should be conceived as a complement rather than a substitute for cognitive effort. Educational and professional frameworks must foster deliberate, reflective use of AI alongside strategies that cultivate critical thinking, autonomous problem solving, and metacognitive awareness (Luckin & Holmes, 2016). In mental health contexts, maintaining equilibrium between technological support and psychological autonomy remains essential, as overreliance on assistive technologies may erode self-efficacy and heighten anxiety when such tools are unavailable (Montag & Walla, 2016).

At a broader level, these results highlight the urgency of developing ethical and reflective models of AI integration, models that recognize the human mind not as a system to be replaced, but as a dynamic entity whose adaptability, creativity, and autonomy must be protected. Longitudinal and neurocognitive research is needed to explore whether sustained AI exposure influences neural plasticity, attentional control, or the distribution of cognitive labor (Green et al., 2012). Ultimately, the relevance of this study lies in reframing the dialogue between humans and intelligent systems: the challenge is not to resist AI, but to ensure that its incorporation preserves the core human capacities, understanding, autonomy, and meaning, that define intelligence itself.

## Supplementary material

**Programmed Activities (available online):** links to the Canva-based interactive tasks and Tower of Hanoi simulation used in the intervention.

- **Day 1: Crossword Puzzle**
  https://www.canva.com/design/DAF-H_NP_jM/CZ_NxSEtrlPuuBYKt_jV_Q/edit?utm_content=DAF-H_NP_jMyutm_campaign=designshareyutm_medium=link2yutm_source=sharebutton
- **Day 2: Problem-Solving Tasks**
  https://www.canva.com/design/DAF-Iu-4jTk/NDw5EVwqpFzq31xVz238hA/edit?utm_content=DAF-Iu-4jTkyutm_campaign=designshareyutm_medium=link2yutm_source=sharebutton
- **Day 3: Word-Guessing Game**
  https://www.canva.com/design/DAF-ICF7Kqw/maXvhfaKwfXp-QGJWLvAdw/edit?utm_content=DAF-ICF7Kqwyutm_campaign=designshareyutm_medium=link2yutm_source=sharebutton
- **Day 4: Lateral Thinking**
  https://www.canva.com/design/DAF-I19Li5k/fL3Go4i1I7_nhVpjiNIUCQ/edit?utm_content=DAF-I19Li5kyutm_campaign=designshareyutm_medium=link2yutm_source=sharebutton
- **Day 5: Trivia**
  https://www.canva.com/design/DAF-ILEOmKM/-

XdObWFPepc_PEpBtlEfcA/edit?utm_content=DAF-ILEOmKMyutm_campaign=designshareyutm_medium=link2yutm_source=sharebutton

- **Day 6: Tower of Hanoi**

  https://www.mathsisfun.com/games/towerofhanoi.html

- **Day 7: Reading Comprehension**

  https://www.canva.com/design/DAF-IRm-zDc/OHRXvcytKxO8gNTGIUVXfQ/edit?utm_content=DAF-IRm-zDcyutm_campaign=designshareyutm_medium=link2yutm_source=sharebutton

- **Day 8: Brainstorming**

  https://www.canva.com/design/DAF-I_8vkow/2VVqfQgG7dlUPZlVKL6a0w/edit?utm_content=DAF-I_8vkowyutm_campaign=designshareyutm_medium=link2yutm_source=sharebutton